\documentclass{aastex}
\usepackage{emulateapj5}

\input{epsf}

\usepackage{times}

\newcommand{\flux}{\,{\rm ph}\,{\rm cm}^{-2}\,{\rm s}^{-1}}
\newcommand{\keV}{\,{\rm keV}}
\newcommand{\lum}{\,{\rm erg}\,{\rm s}^{-1}}
\newcommand{\kpc}{\,{\rm kpc}}
\newcommand{\dM}{\dot{M}}
\newcommand{\LE}{L_{\rm E}}
\newcommand{\grad}{{\rm o}}
\newcommand{\rms}{{\rm rms}}
\newcommand{\hr}{{\rm hr}}
\newcommand{\fhr}{{\rm fhr}}

\slugcomment{The Astrophysical Journal, in press}

\lefthead{STERN, BELOBORODOV, \& POUTANEN}
\righthead{BIZARRE HARD X-RAY OUTBURSTS OF CYGNUS X-1}

\begin{document}

\title{Bizarre Hard X-ray Outbursts of Cygnus X-1}

\author{Boris E. Stern,\altaffilmark{1,2} Andrei M. Beloborodov,   
\altaffilmark{1} and Juri Poutanen}

\affil{Stockholm Observatory, SE-133 36, Saltsj{\"o}baden, Sweden; 
\{stern,andrei,juri@astro.su.se\}}

\altaffiltext{1}{Also at Astro-Space Center of Lebedev Physical Institute,
Profsoyuznaya 84/32, 117810 Moscow, Russia}

\altaffiltext{2}{Also at Institute for Nuclear Research, Russian Academy of 
Sciences, 117312 Moscow, Russia}


\begin{abstract}
A very high  activity  of Cygnus X-1 on 1999 April  19-21 was  recorded  by
BATSE Large Area Detectors onboard the {\it Compton  Gamma-Ray  Observatory
(CGRO)}.  The peak  luminosity  was one order of magnitude  higher than the
normal  luminosity of Cyg X-1.  This fact can be critical for models of the
hard state of Cyg X-1.  The  longest  outburst  lasted  $\sim  1000$~s  and
demonstrated  very unusual  temporal and spectral  behavior which indicates
the presence of two emission  components.  One component is relatively soft
(with a cutoff below $\sim 100\keV$) and highly variable, and the other one
is hard (extending above 100 keV), with much slower variability.
\end{abstract}



\section{Introduction}

Cyg X-1 is a persistent  X-ray  source  believed to be powered by accretion
onto a black  hole  from a  massive  companion.  Most of the  time  Cyg~X-1
spends in the hard state and  sometimes it switches to the soft state (see,
e.g., Liang \& Nolan 1984; Zhang et al.  1997; Poutanen 1998;  Gierli\'nski
et al.  1999).  The X-ray luminosity  above $\sim 2\keV$ is estimated to be
about  $3\times  10^{37}\lum$  (Gierli\'nski  et al.  1997),  assuming  the
distance to the source, $D=2 \kpc$  (Massey,  Johnson, \&  DeGioia-Eastwood
1995; Malysheva 1997).

Variability  of Cyg X-1 was  extensively  studied with various  instruments
(see, e.g., Ling et al.  1987;  Gilfanov et al.  1995; Phlips et al.  1996;
Kuznetsov et al.  1997; Paciesas et al.  1997; Wen et al.  1999;  Brocksopp
et al.  1999; Ba{\l}uci\'nska-Church et al.  2000).  These studies show the
stability  of the hard state of  Cyg~X-1.  The photon  flux above 50 keV is
normally $\sim 0.1\flux$ with variations by a factor~2.

In this  Letter, we analyze the period of the  unusually  strong hard X-ray
activity of Cyg X-1 on 1999 April~21.  Two strong  outbursts  activated the
BATSE/{\it  CGRO}  (Fishman et al.  1989) onboard  trigger.  We analyze the
temporal  structure of the outbursts in the four LAD energy  channels 1--4,
estimate the peak luminosity of Cyg X-1, and discuss possible  implications
for theoretical models of accretion in this object.


\section{The Outbursts}

The unusual  activity of Cyg X-1 can be traced back to 1999  April~19  (TJD
11287).  Two events  recorded  on  April~19  were  found in the BATSE  data
during  search for  non-triggered  gamma-ray  bursts  (GRBs)  
(Stern et al. 1999, 2000a).  The best fit locations were within  
$4^\grad-5^\grad$  from Cyg X-1
($1\sigma$  errors  exceeded  $10^\grad$  for  those  events).  We use  the
location  procedure  described in Stern et al.  (2000b)  which is similar to
that used for GRB location by Pendleton et al.  (1999).  The estimated peak
fluxes above 50~keV were about 0.3 and $0.5\flux$ which is high as compared
with the  normal  flux from Cyg X-1.  On  April~20,  BATSE  detectors  were
triggered by another event with right ascension,  $\alpha=305\fdg 1$, and
declination, $\delta=26\fdg 2$ (BATSE estimate, note also that this event
is  identified  as a GRB in the BATSE data base).  Again, the  location was
close to Cyg X-1 (slightly  beyond  1$\sigma$  error  circle).\footnote{The
coordinates of Cyg X-1 are $\alpha=299\fdg 5$  and  $\delta=35\fdg 2$.}
The peak flux of this event was about $0.5 \flux$.

On April~21, the two brightest outbursts  occurred, with the interval $\sim
2.5$  hours.  Between  and after the two  outbursts,  Cyg X-1  demonstrated
nothing  unusual,  being in the hard  state  with  normal  luminosity.  The
summary of all the five events is given in Table~1.

We will concentrate on the two brightest events on April~21 (two last lines
in Table~1).  The  locations of both events  coincide  with the location of
Cyg X-1 with  2$^\grad$  accuracy.  In  principle,  it might be  accidental
projections  of GRBs on Cyg X-1.  The  probability  that one of $\sim 10^3$
detected GRBs with comparable  brightness will appear within 2$^\grad$ from
Cyg X-1 is about 0.3.  However, the  probability  of appearance of two such
GRBs  from the same  location  within  2.5  hours is low,  $\sim  10^{-5}$.
Besides, the long  duration of the events,  $\sim  10^3$~s, is very unusual
for GRBs.  One concludes that the events are outbursts of Cyg X-1.

\footnotesize
\begin{center}
{\sc TABLE 1\\Summary of Outbursts on 1999 April 19 - 21}
\vskip 4pt
\begin{tabular}{cccccr}
\hline
\hline
ST2000$^{\rm a}$ & \#$^{\rm b}$ & time$^{\rm c}$, s & 
 $\alpha^{\rm d}$ & $\delta^{\rm d}$ & $1\sigma^{\rm d}$\\  
\hline
11287a & \dots & 16816 & 297.8 & 38.2 & 18.4   \\
11287b & \dots & 18071 & 296.6 & 38.0 & 12.9   \\
11288e & 7523  & 83437 & 305.6 & 27.6 &  6.0   \\
11289c & 7524  & 54530 & 299.9 & 34.4 &  2.4   \\
11289e & 7525  & 63100 & 298.4 & 35.2 &  1.2   \\
\hline
\end{tabular}
\end{center}
{
$^{\rm a}$ Name in the catalog of Stern et al. (1999),
consisting of TJD and a letter. \\
$^{\rm b}$ BATSE trigger number. \\
$^{\rm c}$ Time of the day in seconds. \\
$^{\rm d}$ The best fit location and its $1\sigma$ error (deg). 
}
\normalsize

\subsection{Outburst 11289c}

The first  outburst on  April~21  has  similar  light  curves in LAD energy
channels  1--3 (see top panel in Fig.~1).  The signal in  channel~4  (above
300~keV)  was low,  practically  undetectable.  The  signal  was cut off by
Earth  occultation at $\approx  55.23$~ks.  We model the background  during
the outburst  using a linear fit  extrapolating  the  background  after the
occultation.

In order to get the photon and energy  fluxes, we fit the signal count rate
in the three energy channels using the detector  response  matrix  computed
with the code of  Pendleton  (see  Pendleton  et al.  1999  and  references
therein).  An  exponentially  cutoff  power-law  is  assumed  as a spectral
hypothesis.  The photon and energy fluxes are then obtained by  integrating
the  best  fitting  spectrum  in the  proper  energy  band.  The  error  is
estimated to be about 15\%. (Note that the overall systematic error in the
BATSE flux normalization may be about 20\%,  see, e.g., Much et al. 1996.)

\medskip
\centerline{\epsfxsize=7.5cm \epsfbox{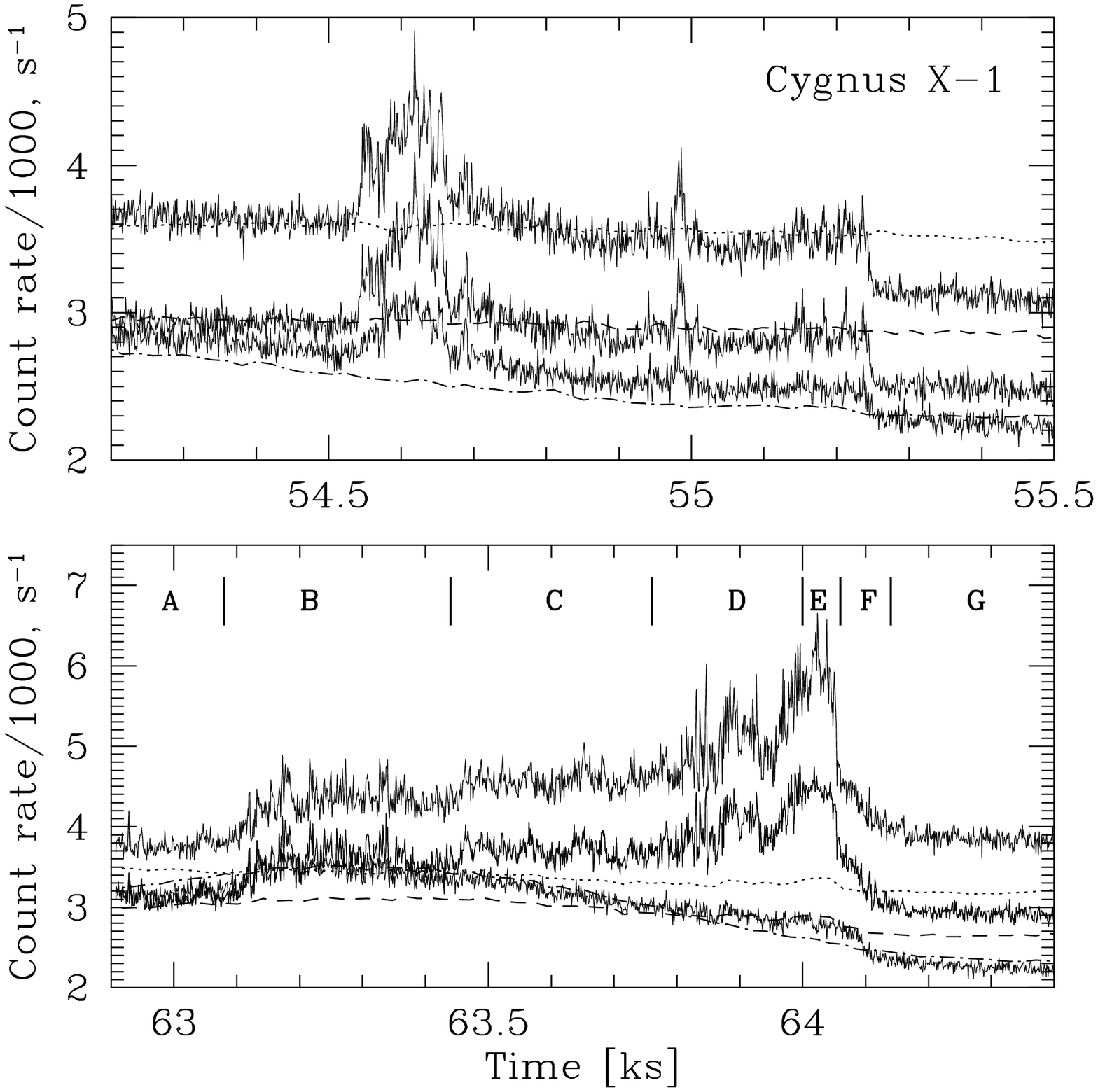}}  
\figcaption{
Count rate during the two outbursts of Cyg X-1 on 1999 April 21 (TJD 11289)
in the three LAD energy  channels, 1--3.  The count rate is summed over two
detectors  closest to the line of sight to Cyg~X-1.  Count rates are higher
in  softer  channels.  Dotted,  dashed  and  dot-dashed   curves  show  the
background in channels 1, 2, and 3, respectively,  as seen by two detectors
looking away from Cyg~X-1.
\label{fig:lc}}
\footnotesize
\begin{center}
{\sc TABLE 2\\
Peak Fluxes and Luminosities}
\vskip 4pt
\begin{tabular}{ccccc}
\hline
\hline
 Outburst          & $F_{>50}$$^{\rm a}$   &  $F_{>30}$$^{\rm b}$ &  
 $L_{>50}$$^{\rm c}$ & $L_{>30}$$^{\rm d}$  \\
\hline
11289c & 1.1 & 2.4 & $0.75\times10^{38}$ & $1.2\times10^{38}$  \\
11289e & 1.3 & 3.9 & $0.8\times10^{38}$  & $1.6\times10^{38}$  \\
\hline
\end{tabular}
\end{center}
\setcounter{table}{2}
{
$^{\rm a}$ Peak flux ($\flux$) above 50~keV. \\
$^{\rm b}$ Peak flux above 30~keV.\\
$^{\rm c}$ Peak luminosity ($\lum$) above 50~keV 
(assuming distance $D=2$~kpc). \\
$^{\rm d}$ Peak luminosity above 30~keV.
} 
\medskip
\normalsize

The peak  flux in 50--300 keV band  (channels  2--3) is about  $1.1\flux$
which is $\sim10$ times higher than the normal flux from Cyg X-1.  The flux
averaged over hundred  seconds  (54.55-54.65~ks  interval) is  $0.75\flux$.
The peak  luminosity  above 30~keV exceeds  $10^{38}\lum$  (see Table~2 for
details).  Note  that  the  strong  increase  in  the  luminosity  was  not
accompanied  by crucial  changes in the  spectrum.  At least, the  hardness
ratios during the outburst  stayed  similar to the normal hard state of Cyg
X-1 (see \S~3).

\subsection{Outburst 11289e}

The second  outburst on April~21 was  detected  $\sim$ 8~ks after the first
outburst (see bottom panel in  Fig.~\ref{fig:lc}).  Unfortunately, there is
a gap in the  data  at  the  time  when  Cyg~X-1  rose  above  the  horizon
($\sim$62.8~ks).  The data records  start at $\approx  62.9$~ks.  Note that
there was no  occultation of Cyg X-1 in the end of the outburst.  The sharp
fall off in the light curves in each channel is the  intrinsic  behavior of
the source.  It allows one to estimate the peak flux and  luminosity of the
outburst which we give in Table~2.  The corresponding luminosities averaged
over interval E (see Fig.~1) are 20\% smaller than the peak values.

The ionospheric  background  showed smooth latitude  variations  during the
outburst (see bottom panel of Fig.~1) on the time-scale of order 1~ks.  The
background  level is different in  different  detectors,  therefore,  it is
difficult to extract  exactly the signal seen in the  detectors  looking at
Cyg~X-1.  The signal uncertainty caused by the background  variations is of
order 20\%. 

One can get a good  location fit for the signal by analyzing  the  variable
part of the light curve.  At a given time  interval, we fit the light curve
by a  linear  function  and  take it as a  ``reference''  level.  (It  thus
includes both the  background  and the linear part of the true signal.)  We
then  take the  difference  between  the  count  rate (at  this or  another
interval)  and the  reference  level as a  ``signal''.  The  location  fits
demonstrate  that  almost all  variable  part of the signal  comes from the
direction of Cyg X-1 with $\sim 5^\grad$ accuracy at all time intervals A-F
marked in Figure~1  (see  Table~3).  The  residual  signal can be fitted by
200~s -- 400~s long linear  fragments  with a good  $\chi^2$  (lines 1-3 in
Table~3).  This  means  that no  fraction  of the  variable  signal  can be
mimicked by  magnetospheric phenomena  like  particle  precipitation  which
either  produces a diffuse flux of photons or can be a localized  source at
some distance  from the  spacecraft.  In the first case the location fit is
so bad that the residual  $\chi^2$ is comparable to the initial one, and in
the second case one observes a fast change in the  direction  of the source
due to the satellite  motion.  In our case, the satellite moved $\sim 6000$
km between  intervals B and E, while the  source's  position on the sky did
not change.

\footnotesize
\begin{center}
{\sc TABLE 3\\
Location Fits for Different Time Intervals}
\vskip 4pt
\begin{tabular}{cccccrc}
\hline
\hline  
Signal$^{\rm a}$ & Reference$^{\rm b}$  & Channel & $\alpha^{\rm c}$  & 
$\delta^{\rm c}$ & $1\sigma^{\rm c}$ & $\chi^2_0/\chi^2_r$ $^{\rm d}$ \\
\hline
B   & B   & 2+3 & 302.1 & 37.1 &  6.6  & 11.4 / 0.93 \\
D+E & D+E & 1   & 302.3 & 38.4 &  2.0  & 34.0 / 1.13 \\
D+E & D+E & 2+3 & 301.3 & 40.2 &  6.0  & 16.9 / 1.05 \\
D+E & G   & 2+3 & 298.4 & 35.2 &  1.2  & 447  / 10.8  \\
  E & E   & 1   & 303.9 & 40.1 &  4.9  & 15.3 /  0.90  \\
E+F & G   & 3   & 298.8 & 33.5 &  2.1  & 64.4 / 2.35 \\
F   & G   & 2+3 & 302.4 & 40.3 &  4.1  & 51.0 / 1.66 \\
\hline
\end{tabular}
\end{center}
\setcounter{table}{3}
{
$^{\rm a}$  The time interval in which the location fit is done for the residuals
      after subtracting a linear (``reference'') function from the count rate. \\
$^{\rm b}$  The  interval at  which  the linear reference function was fitted. \\
$^{\rm c}$  The best fit location and its $1\sigma$ error (deg). \\
$^{\rm d}$  $\chi_0^2$ and $\chi^2_r$  are $\chi^2$ per dof for the count rate
relative to the reference level and for the residual signal 
after the location fit, respectively.
}
\medskip 
\normalsize

While the first  April~21  outburst is unusual only for its  strength,  the
second one is even more  surprising.  The  behavior  of the light  curve in
different energy channels suggests the presence of two independent emission
components.  The first (highly variable) component  dominates the signal in
channels  1 and  2,  and  the  second  component  (with  low  amplitude  of
variability)   dominates  the  signal  in  channel~3.  The  soft  component
terminated at 64.05~ks and then one observes only the hard component in the
three  channels.  Unfortunately,  the spectral data for this  outburst were
lost,  which did not allow us to confirm  the  two-components  by  spectral
analysis.

Since we cannot exactly  separate the signal from the  background, we study
the  outburst  using the strong  non-Poisson  variability  produced  by the
signal in the count  rate.  In each  time  interval,  A-F, we  compute  the
root-mean-square  (rms) of the signal variations,  $S_i$, around the linear
fits ($i=1,2,3$ is the channel number).  We then compute the  ``fluctuation
hardness ratio'', fhr, that is the ratio of the rms in different  channels.
The  rms  (with   subtracted   Poisson   component)  and   $\fhr_{21}\equiv
\rms_2/\rms_1$  are  given  in  Table~4.  Assuming  the  rms to be  roughly
proportional  to the  average  level  of the  signal,  the fhr may  give an
estimate for the true hardness  ratio.  The decrease of the  $\fhr_{21}$ as
the outburst progresses  indicates that the variable component gets softer.

\footnotesize
\begin{center}
{\sc TABLE 4\\
Characteristics of the Count Rate Variability}
\vskip 4pt
\begin{tabular}{cccccc}
\hline
\hline
 Interval           & $\rms_1$   &  $\rms_2$  &  
 $\rms_3$ & $c_{23}$  & $\fhr_{21}$  \\
\hline
A & 97  & 109 & 70  & 0.63 $\pm$ 0.09 & 1.12 $\pm$ 0.06 \\
B & 300 & 316 & 169 & 0.92 $\pm$ 0.02 & 1.05 $\pm$ 0.02  \\
C & 211 & 213 & 96  & 0.77 $\pm$ 0.05 & 1.01 $\pm$ 0.03  \\
D & 499 & 370 & 100 & 0.71 $\pm$ 0.04 & 0.74 $\pm$ 0.01   \\
E & 568 & 286 & 90  & 0.15 $\pm$ 0.07 & 0.50 $\pm$ 0.02  \\
F & 109 & 196 & 106 & 0.92 $\pm$ 0.07 & 1.32 $\pm$ 0.10   \\ 
\hline
\end{tabular}
\end{center}
\setcounter{table}{4}
{
The $1 \sigma$ errors of $c_{23}$ and $\fhr_{21}$
are of Poisson nature; they are estimated with the bootstrap method.
The measured hardness ratios 
are not affected by the background uncertainty because they 
are based on the signal dispersion in a narrow time interval where the 
background is almost linear.   
} 
\medskip
\normalsize

We then compute the cross-correlation coefficient between channels 2 and 3,
$c_{23}\equiv  (\langle S_2 \cdot S_3 \rangle - \langle  S_2\rangle \langle
S_3   \rangle)/(\rms_2   \rms_3$)  (see  Table~4).  The   cross-correlation
decreases  during the  outbursts  and becomes  very low in interval  E.  It
confirms that the soft component gets so soft that it practically  does not
contribute   to   the   signal   in   channel~3.   In   interval   F,   the
cross-correlation   is  high,   confirming  that  the  soft  component  has
disappeared  and we see only the second  (hard)  component in all the three
channels.  The  luminosities of the soft and hard components can be roughly
estimated   from  the   step-like   cut  offs  at  64.05~ks   and  64.1~ks,
respectively.  The average  luminosity of Cyg~X-1 in time  interval E above
50~keV  (30~keV)  is  $L_{>50}\sim  2.5  \times10^{37}\lum$   ($L_{>30}\sim
7.0\times10^{37}\lum$)    in   the   soft   component   and   $L_{>50}\sim
4.0\times10^{37}\lum$   ($L_{>30}\sim5.5\times10^{37}\lum$)   in  the  hard
component.

\section{Comparison with the Normal Hard State of Cyg X-1}

The normal hard state of Cyg X-1 was studied  using the Earth  occultations
which  occur on each orbit of the {\it CGRO}.  The step in the light  curve
at the moment of occultation shows the amplitude of the signal from Cyg
X-1 in each energy channel.  The study of Cyg X-1 with this method was done
by Ling et al.  (1997) and Paciesas et al.  (1997).  We performed a similar
analysis for $\sim 60$ occultations and evaluated the flux from Cyg X-1 and
the  hardness ratio $\hr_{32}$ (ratio of the signal count rates 
in channels 3 and 2)  for each  occultation.  The  results  are  shown  in
Figure~2. 
The observed large dispersion in the  hardness  ratio
may be caused by the measurement errors rather than  
intrinsic variability of Cyg~X-1.
To estimate the errors we did a similar analysis of 80 occultations of 
the Crab nebula which is known to be a steady source. We found 
$\hr_{32}=0.66\pm 0.13$ and the count rate $316\pm 110$, where
the errors represent the standard deviations.
The major cause of the errors is the uncertainty in the changing background 
(to have good statistics one has to fit the signal on a relatively long period 
$\sim 100$ s before and after the occultation, and the background curvature 
plays a role).
One should note that the error in the hardness  ratio 
of Cyg X-1 is likely to be smaller than that for the Crab 
since its flux is larger. In the outbursts, these errors are even smaller.
 
\medskip
\centerline{\epsfxsize=7.5cm   \epsfbox{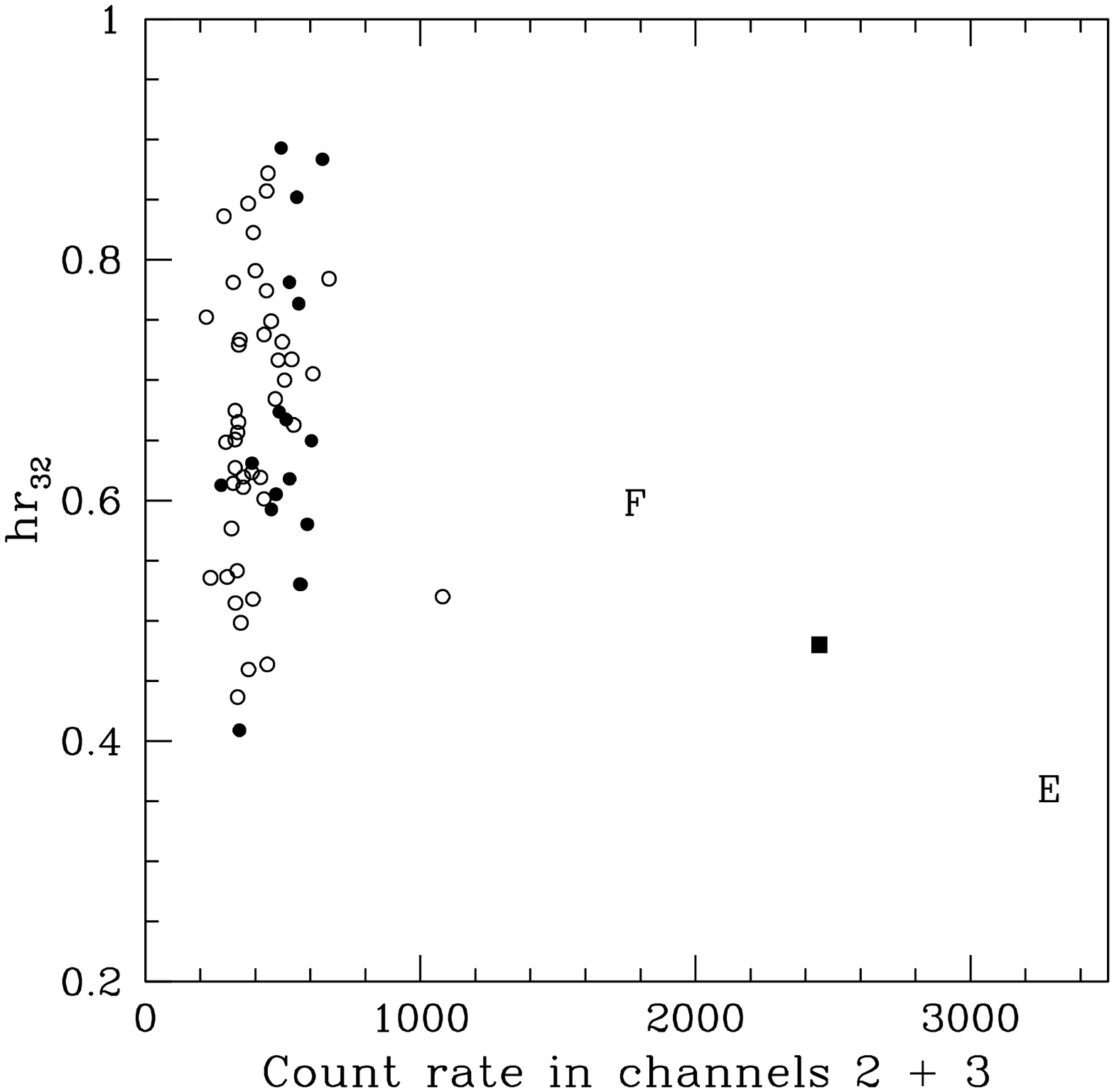}} 
\figcaption{Hardness  ratio,  $\hr_{32}$,  of  the signal count  rates  in
channels 3 and 2 versus  count rate in channels  2+3, as measured 
in all  detectors   (note the difference from Fig.~1 where  
the count rate is given only for two detectors).
{\it Open circles} show a sample of occultations  in the period TJD 
8600--8700; {\it  filled  circles}  show the same in the period TJD 9990--10000.
Cyg X-1 was in the normal hard state with  relatively  high  brightness  in
both periods.  
The dispersion in $\hr_{32}$ may be caused by measurement errors (see the text).
{\it Filled  square}  corresponds  to the first  outburst on
April 21 (between 54.55 and 54.65~ks). 
Labels F and E represent the second
outburst on April~21  in time intervals F and E,  respectively. (The signal
was  roughly  estimated  by  subtracting  the  level of the  count  rate in
interval G.)
}
\medskip

For  comparison  we also  show  the  estimates  of the  flux and
hardness   ratio  for  the  two  April~21   outbursts.   
Note that outburst 11289c  and   interval F of outburst 11289e 
have hardness ratios similar to the normal hard state. 
Interval E of outburst 11289e  is significantly softer.

The presence of two emission  components  (soft, highly variable, and hard,
with lower variability) in outburst 11289e is intriguing.  Are
they  present in the  normal  state of  Cyg~X-1? In the sample of 63 
occultations we found that the fractional rms of the count rate 
(i.e. the ratio of the rms to the mean count rate) in channel 2
was higher than that in channel 3 in 47 cases.  
This indicates higher variability in channel 
2 (the probability of such an accidental excess is $0.5 \cdot 10^{-4}$). 
The decreasing  of the  variability  with energy was also  observed in the
2--40 keV band (Nowak et al.  1999;  Revnivtsev et al.  2000).  These facts
are  consistent   with  the  presence  of  a  hard  component  with  a  low
variability though this interpretation  is not  unique.  Note also that
Gierli\'nski  et al.  (1997)  got  the  best  fit to the  broad-band  X-ray
spectrum of Cyg~X-1 with two thermal Comptonization components of different
temperatures.


\section{Discussion}

The hard  state of  Cyg~X-1  has been a puzzle  since  its  discovery.  The
standard  accretion disk model (Shakura \& Sunyaev 1973) was not able to
explain  the  X-ray  spectrum,  and two  modifications  of the  model  were
suggested:  a  two-temperature  hot  disk  and an  active  corona  atop the
standard disk (see Beloborodov 1999a for a recent review).

The advective  hot-disk models (e.g. Esin et al. 1998)
are consistent with the observed  spectrum
and luminosity of Cyg X-1 if the accretion rate has a specific value $\dM
\approx  \dM_{\rm  max}\approx$ $10\alpha^2L_E/c^2$  and  the  viscosity
parameter  $\alpha \approx  0.2-0.3$.  Here $\LE=$ $4\pi c G M m_p/ \sigma_T=$
$1.3 \times  10^{39}$ $(M/10M_{\odot})\lum$  is the Eddington  luminosity.  
Small   variations  in  the   accretion   rate,
$\Delta\dM/  \dM \sim 10\%$, were predicted to destroy the hard state (Esin
et al.  1998; but see also Zdziarski 1998).  By contrast, the luminosity of
Cyg X-1 is known to vary by a factor of two without substantial  changes in
the  spectral  shape  (e.g.,  Paciesas  et al.  1997;  Gierli\'nski  et al.
1997).  Such fluctuations already challenged the model, and the
hard outbursts analyzed in this Letter are even more difficult to explain.
The model would work only assuming a specific  dependence  $\alpha \propto 
\dM^{1/2}$ that   keeps $\dM \approx \dM_{\rm max}$; $\alpha$ then should
be about unity at the peak of the outburst.  

In the context of the  disk-corona  model, the outbursts can be interperted
as an enhanced coronal  activity of the accretion disk.  In this model, the
X-ray spectral slope is controlled by one  parameter,  the feedback  factor
due to X-ray reprocessing by the disk.  The hard-state  spectrum of Cyg~X-1
is well explained if the coronal  plasma is ejected away from the disk with
a mildly relativistic velocity $\beta=v/c\sim 0.3$ (Beloborodov 1999b;
Malzac, Beloborodov, \& Poutanen 2001).
Alternatively,  the observed  emission  may be produced by a static  corona
atop a strongly  ionized disk (e.g., Ross, Fabian, \& Young 1999; Nayakshin
1999).  The  corona      becomes   $e^{\pm}$--dominated   at  high
luminosities and   its   temperature decreases (e.g., Svensson
1984; Stern et al.  1995; Poutanen \& Svensson 1996).  Pair creation may 
cause the shift of the spectral break in outburst 11289e to smaller 
energies.

Cyg~X-1 is a massive X-ray binary and it may be fed mainly by the donor wind. 
Then the
pattern of accretion  can change completely compared to the standard
viscous $\alpha-$disk or its modifications.  The captured wind matter has a
low angular momentum, just about critical for disk formation
(Illarionov  \&  Sunyaev  1975).  Under  such conditions, a small-scale 
inviscid disk   forms, which accretes super-sonically
(Beloborodov  \&  Illarionov  2001).  
The disk forms in the ring-like caustic of the accretion flow  
where energy is  liberated in inelastic collision of gas streams. 
A Comptonized  power-law spectrum is
then emitted with a standard break at $\sim 100$~keV, and it has appearance
of a normal hard state of Cyg~X-1.

The two-component emission in outburst 11289e probably requires a two-zone
emission model.  For instance, the soft component may be associated  with a
variable  coronal  emission atop the disk, and the hard  component  with an
inner relativistic jet. A corona-jet  model was recently
proposed by  Brocksopp  et al.  (1999)  based on  correlations  between the
X-rays and the radio emission observed in Cyg X-1.
The April 1999 outbursts
were, however, too short to have a significant  impact on the flux
in radio and no substantial changes were  detected (R.~Fender and G.~Pooley,
private communication).  In the context of the wind-fed  accretion model,
the  two-component  emission can naturally appear if viscous accretion disk 
and wind-disk collision are both operating in the source.

Concluding, the 1999 April 21  outbursts are the brightest  events detected
from Cyg X-1 by BATSE  during 9 years  of its  operation.  The  hard  X-ray
luminosity  above  30~keV  was in excess of  $10^{38}\lum$.  Unfortunately,
there are no simultaneous  observations in the soft X-rays, so that one can
only guess what was the total luminosity. The large luminosity and 
unusual spectral behavior challenge the theoretical models of accretion
in Cyg X-1.   

\acknowledgments

We thank Rob Fender, Guy Pooley, and Jerry Bonnell for useful  discussions.
We are grateful to Rob Preece and Geoff  Pendleton for  providing  the code
for computing the detector  response matrix and an anonymous referee
for  useful comments that significantly improved  the paper.  
This research has made use of
data obtained through the High Energy Astrophysics Science Archive Research
Center Online Service,  provided by the  NASA/Goddard  Space Flight Center.
This work was supported by the Swedish Natural  Science  Research  Council,
the Swedish  Royal  Academy of Sciences,  the  Wenner-Gren  Foundation  for
Scientific  Research, the  Anna-Greta  and Holger  Crafoord  Fund, and RFBR
grant 00-02-16135.


\end{document}